\documentclass[lettersize,journal]{IEEEtran}
\IEEEoverridecommandlockouts

\usepackage{cite}
\usepackage{amsmath,amssymb,amsfonts}
\usepackage{graphicx}
\usepackage{textcomp}
\usepackage{xcolor}
\usepackage{subfigure}
\usepackage{multirow}
\usepackage{booktabs}
\usepackage{hyperref}
\usepackage{algorithm}
\usepackage{algorithmic}
\usepackage{marvosym}

\def\BibTeX{{\rm B\kern-.05em{\sc i\kern-.025em b}\kern-.08em
    T\kern-.1667em\lower.7ex\hbox{E}\kern-.125emX}}
\begin{document}

\markboth{IEEE TRANSACTIONS ON MOBILE COMPUTING}%
{Shell \MakeLowercase{\textit{et al.}}: A Sample Article Using IEEEtran.cls for IEEE Journals}

\makeatletter
\newcommand{\rmnum}[1]{\romannumeral #1}
\newcommand{\Rmnum}[1]{\expandafter\@slowromancap\romannumeral #1@}
\makeatother

\title{\LARGE \bf When Motion Learns to Listen: Diffusion-Prior Lyapunov Actor–Critic Framework with LLM Guidance for Stable and Robust AUV Control in Underwater Tasks
}

\author{Jingzehua Xu$^{1,\dag}$, \IEEEmembership{Student Member, IEEE}, Weiyi Liu$^{1,\dag}$,  \IEEEmembership{Student Member, IEEE},\\Weihang Zhang$^{2,\dag}$, Zhuofan Xi$^{3}$, Guanwen Xie$^{1}$, \IEEEmembership{Student Member, IEEE},\\Shuai Zhang$^{4}$, \IEEEmembership{Member, IEEE}, and Yi Li$^{1,}\textsuperscript{\Letter}$, \IEEEmembership{Member, IEEE}
\thanks{This article has be presented in part at the IEEE/RSJ International Conference on Intelligent Robots and Systems (IROS), Hangzhou, China, in October
2025.}%
\thanks{$^{1}$J. Xu, W. Liu, G. Xie and Y. Li are with Tsinghua Shenzhen International Graduate School, Tsinghua University, Shenzhen, 518055, China, and with Department of Engineering, University of Cambridge, CB2 1PZ, United Kingdom. Email: 19955778426@163.com, \{liuwy24, xgw24\}@mails.tsinghua.edu.cn, liyi@sz.tsinghua.edu.cn.}%
\thanks{$^{2}$W. Zhang is with Department of Electrical and Computer Engineering, Duke University, Durham, NC 27708, USA. E-mail: weihang.zhang@duke.edu.}
\thanks{$^{3}$Z. Xi is with Stanford University, Stanford, CA 94305, USA. Email: zfxi624@gmail.com.}
\thanks{$^{4}$S. Zhang is with Department of Data Science, New Jersey Institute of Technology, NJ 07102, USA. Email: sz457@njit.edu.}
\thanks{$^\dagger$ These authors contributed equally to this work.}
\thanks{$\textsuperscript{\Letter}$ Corresponding author.}
}

\maketitle

\begin{abstract}
Autonomous Underwater Vehicles (AUVs) are indispensable for marine exploration; yet, their control is hindered by nonlinear hydrodynamics, time-varying disturbances, and localization uncertainty. Traditional controllers provide only limited adaptability, while Reinforcement Learning (RL), though promising, suffers from sample inefficiency, weak long-term planning, and lacks stability guarantees, leading to unreliable behavior. To address these challenges, we propose a diffusion-prior Lyapunov actor–critic framework that unifies exploration, stability, and semantic adaptability. Specifically, a diffusion model generates smooth, multimodal, and disturbance-resilient candidate actions; a Lyapunov critic further imposes dual constraints that ensure stability; and a Large Language Model (LLM)-driven outer loop adaptively selects and refines Lyapunov functions based on task semantics and training feedback. This “generation–filtering–optimization” mechanism not only enhances sample efficiency and planning capability but also aligns stability guarantees with diverse mission requirements in the multi-objective optimization task. Extensive simulations under complex ocean dynamics demonstrate that the proposed framework achieves more accurate trajectory tracking, higher task completion rates, improved energy efficiency, faster convergence, and improved robustness compared with conventional RL and diffusion-augmented baselines.
\end{abstract}

    \begin{IEEEkeywords}
    Diffusion Model, Reinforcement Learning, Lyapunov Function, Autonomous Underwater Vehicle, Large Language Model, Robust Control, Underwater Tasks. 
    \end{IEEEkeywords}
\section{INTRODUCTION}
\label{sec:intro}
Autonomous Underwater Vehicles (AUVs) have been widely applied in missions such as deep-sea mapping, ecological monitoring, and underwater infrastructure inspection \cite{1,2,3}. However, operating in extreme ocean environments poses three major challenges to their control: first, the complex nonlinear hydrodynamics and thruster dynamics significantly increase modeling difficulty \cite{4}; second, time-varying disturbances such as currents and waves continuously affect motion stability \cite{5}; and third, sensor noise and accuracy limitations cause localization uncertainty, which propagates as state estimation errors and increases the difficulty of maintaining stable and robust control. \cite{6}. These factors jointly impose stringent requirements on AUV control, necessitating an adaptive balance among key metrics such as trajectory tracking accuracy, energy efficiency, and operational reliability in the unstructured environment \cite{23}.

To address these issues, traditional control methods have offered partial solutions; however, their capabilities remain limited——Proportional-Integral-Derivative (PID) controllers are simple and general, but their fixed-gain design cannot adapt to highly variable ocean conditions \cite{7,23}. Model Predictive Control (MPC) provides strong optimization; however, its computational complexity grows quickly with the prediction horizon, making real-time use infeasible for resource-constrained AUVs \cite{8,24}. Sliding Mode Control (SMC) is robust against bounded disturbances, but its chattering effect increases energy consumption and mechanical wear \cite{9,26}. More critically, these approaches rely on prior models or lack the capacity for adaptive learning or long-horizon optimization. Hence, a new approach is needed that supports continuous learning and adaptation in complex, dynamic environments \cite{27}.

\begin{figure*}[!t]
        \centering
        \includegraphics[width=0.99\linewidth]{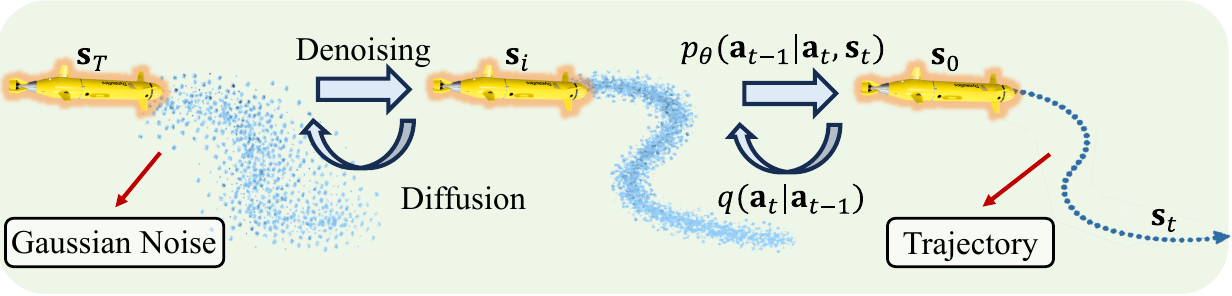}
        \caption{\small \textbf{Illustration of the diffusion model principle}. Through the forward process $q(\mathbf{a}_t \mid \mathbf{a}_{t-1})$ and the reverse process $p_\theta(\mathbf{a}_{t-1} \mid \mathbf{a}_t, \mathbf{s}_t)$, the model progressively refines noisy actions into a coherent control sequence, ultimately generating a smooth AUV trajectory from $\mathbf{s}_0$ to the current state $\mathbf{s}_t$.}
        \label{fig_1}
\end{figure*}

Against this backdrop, Reinforcement Learning (RL) offers a promising pathway. By interacting with the environment, RL can autonomously learn complex control policies \cite{10} and has already demonstrated potential in tasks such as AUV path following and station keeping \cite{11,12,29}. Nevertheless, the application of RL in the AUV domain is constrained by two critical bottlenecks: on the one hand, the exploration process in high-dimensional state spaces is often sample-inefficient, leading to high training costs \cite{13,28}; on the other hand, RL typically lacks long-term planning capability, making it difficult to account for global mission objectives in long-duration deployments \cite{14,27}. These limitations hinder its effectiveness in real-world AUV control \cite{25}.

To overcome these deficiencies, the recently emerging diffusion model provides a novel perspective. As a powerful generative framework, it excels in modeling complex distributions and generating temporal sequences \cite{15}. As depicted in Fig. 1, through iterative denoising, the diffusion model can produce smooth, physically feasible, and disturbance-resilient trajectories \cite{16}. When combined with RL, the diffusion model addresses several critical shortcomings: it can generate diverse long-horizon candidate actions, thereby alleviating RL’s limitations in long-term planning \cite{17}; it naturally incorporates physical constraints at the action level, improving feasibility and robustness under dynamic hydrodynamic conditions \cite{18}; and it enables high-quality exploration trajectories, which enhance sample efficiency in sparse-reward settings, accelerate policy convergence, and yield more generalizable strategies \cite{19}. Together, this “Diffusion + RL” synergy offers a promising new direction for robust AUV control.

Despite recent advances, existing methods still lack a unified guarantee of convergence and safety \cite{21}. In complex ocean environments, even if diffusion-generated candidate trajectories are physically feasible, the absence of explicit stability constraints in policy updates may still cause oscillations or divergence during training \cite{22}, and AUVs may perform actions that compromise reliability during exploration. These limitations highlight the need for stronger theoretical foundations to ensure the robust application of “Diffusion + RL” in real-world AUV control settings \cite{Adaptive1}. Moreover, current approaches cannot adaptively adjust stability properties in response to task feedback and changing objectives, further limiting their practicality in dynamic task requirements \cite{Adaptive2}.

Motivated by these challenges, this paper proposes a diffusion-prior Lyapunov actor–critic framework for AUV control that improves stability, robustness, and adaptability. A neural Lyapunov function is embedded in the critic to impose dual constraints during policy updates, ensuring stability while suppressing unreliable behaviors. To address the lack of adaptive adjustment, we introduce a Large Language Model (LLM)-driven outer loop that leverages task descriptions, semantic reasoning, and training feedback to generate and refine Lyapunov functions, enabling constraints to adapt dynamically to diverse requirements. Combined with diffusion-generated candidate actions, the proposed framework establishes a “generation–filtering–optimization” pipeline that unifies performance, stability, and reliability, offering a principled pathway for robust AUV control.

In summary, the main contributions of this paper can be listed as follows:
\begin{itemize}
\item \textbf{Diffusion–Prior Lyapunov Actor–Critic Framework}: We propose a novel actor-critic framework in RL that integrates diffusion priors with Lyapunov-based constraints in policy updates, thereby achieving stability while addressing the absence of convergence guarantees in conventional RL.

\item \textbf{An LLM-Driven Outer-Loop Adaptation Mechanism}: To further enhance adaptability, we introduce an LLM that leverages task semantics and training feedback to generate and refine Lyapunov functions, enabling stability constraints to adjust dynamically across requirements.

\item \textbf{Extensive High-Fidelity Evaluations}: Finally, through comprehensive simulations, we demonstrate the proposed framework’s consistent improvements in more accurate trajectory tracking, higher task completion and energy efficiency, faster convergence, and improved robustness, benchmarked against conventional RL and diffusion-enhanced RL baselines.
\end{itemize}
The remainder of this paper is organized as follows: Section II reviews the prior related work. Section III details the methodology of this work. Section IV describes the environmental simulations and experimental results. Finally, Section V concludes the paper and outlines future work.

\section{RELATED WORK}
This section reviews prior research most relevant to our work, organized into four directions: (A) RL for AUV control, (B) Lyapunov-based RL and stability guarantees, (C) diffusion models for generative control, and (D) LLMs for semantic adaptation in control.

\subsection{Reinforcement Learning for AUV Control}

RL has gained increasing attention in AUV control because of its capability to learn policies directly from interaction, thereby reducing dependence on accurate hydrodynamic models \cite{Yu2025AUVRLsurvey}. For instance, Jiang \textit{et al.} proposed an attention-based meta-RL framework for trajectory tracking under time-varying dynamics, which demonstrated improved adaptability across different ocean conditions \cite{10}. Similarly, Chu \textit{et al.} developed a deep RL-based path planning method for AUVs operating under ocean current disturbances, showing enhanced robustness compared to conventional planners \cite{Chu2022AUVPathRL}. These studies highlight RL’s potential in handling nonlinear dynamics and uncertain environments.  

Despite these successes, RL still suffers from two critical limitations in real-world AUV deployment. First, exploration in high-dimensional continuous state spaces is often sample-inefficient, resulting in prohibitive training costs and slow convergence \cite{depth}. Second, learned policies lack explicit guarantees of stability, which can cause oscillations or unreliable behaviors when faced with strong disturbances or sensor uncertainties \cite{Hsu2019RLReview}. These limitations underscore the necessity of integrating additional mechanisms, such as stability guarantees, into RL frameworks for reliable underwater operation.

\subsection{Lyapunov-Based Reinforcement Learning and Stability Guarantees}

To address the lack of stability guarantees in conventional RL, researchers have increasingly incorporated Lyapunov functions into the learning process, enforcing a monotonic decrease along system trajectories to ensure the asymptotic or even exponential stability of learned policies. Early work by Perkins and Barto introduced Lyapunov-based constraints to safe RL, establishing one of the first theoretical links between RL and stability analysis \cite{Perkins2002LyapunovRL}. Building on this foundation, Westenbroek \textit{et al.} proposed Lyapunov-guided policy optimization in robotic systems, showing improvements in robustness and efficiency for continuous control \cite{Westenbroek2022LyapunovDesign}. Han \textit{et al.} extended the idea to actor–critic structures, where a Lyapunov critic constrains policy updates to guarantee stability while optimizing reward \cite{Han2020ActorCriticStability}, and Zhao \textit{et al.} further integrated Lyapunov and control barrier functions into a unified critic, allowing agents to satisfy both stability and safety requirements in dynamic environments \cite{Du2023LyapunovBarrierRL}.

Although these methods provide a strong theoretical foundation for safe and stable RL, most existing work has been validated in terrestrial robotics or benchmark environments, such as locomotion and grid-world tasks \cite{Stable}. Their application to marine robotics, particularly in AUV control under highly nonlinear hydrodynamic effects and uncertain disturbances, remains largely unexplored. This gap highlights the importance of extending Lyapunov-based RL methods to AUV domains where both adaptability and stability are critical.

\subsection{Diffusion Models for Generative Control}
In parallel with stability-oriented methods, the machine learning community has explored generative models to improve policy expressiveness and exploration. Among them, diffusion models have emerged as a powerful tool for generating smooth, multimodal, and physically feasible trajectories \cite{18}. Unlike conventional noise-injection strategies, they learn structured action distributions and produce long-horizon, disturbance-resilient proposals \cite{Bai2025DiffusionRobotics}. Building on this potential, Zhang \textit{et al.} proposed a spatial–temporal diffusion model for underwater scene reconstruction, enhancing AUV navigation by capturing spatial structure and temporal consistency \cite{22}. Similarly, Guo \textit{et al.} developed an adaptive AUV hunting policy with covert communication, where diffusion priors supported robust planning and low-detectability coordination \cite{Guo}. Safety-aware adaptations such as CoBL-Diffusion further embed Lyapunov and barrier constraints into the denoising process, enabling safer proposals in dynamic settings \cite{Mizuta2024CoBLDiffusion}. Despite these advances, diffusion-enhanced RL for AUV control remains underexplored, and integrating diffusion priors with formal stability guarantees is still an open problem.

\subsection{Large Language Models for Semantic Adaptation in Control}
Beyond generative modeling, recent advances in LLMs offer a complementary perspective by providing semantic adaptability and task-level reasoning \cite{OpenAI2023GPT4}. LLMs have demonstrated remarkable capabilities in semantic reasoning, cross-task adaptation, and natural language-guided decision-making, motivating their adoption in RL and control \cite{Chowdhery2022PaLM}. Recent works have explored the use of LLMs for policy guidance, reward shaping, and interpretable control decisions, showing that natural language can serve as an effective bridge between high-level goals and low-level actions \cite{Zhou2022PromptRL, Honovich2022Instruction}. However, most of these efforts remain focused on instruction-following or reward adaptation, without addressing stability guarantees \cite{LLM}. In contrast, our approach leverages LLMs as an \textit{outer-loop semantic optimizer}, which selects and refines Lyapunov functions based on task descriptions and training feedback. This integration introduces a novel mechanism for semantic adaptability in stability-constrained RL, enabling policies to remain both task-aware and theoretically grounded in challenging underwater environments.

In summary, RL has been applied to AUV control but suffers from sample inefficiency and instability; Lyapunov-based RL provides formal guarantees but has rarely been adapted to AUVs; diffusion models improve exploration and robustness but lack convergence guarantees; and LLMs offer semantic adaptability but have not been linked to stability enforcement. These gaps directly motivate our proposed framework, which unifies diffusion-based exploration, Lyapunov-constrained RL, and LLM-guided semantic adaptation to achieve robust and exponentially stable AUV control.

\begin{figure*}[!t]
        \centering
        \includegraphics[width=0.99\linewidth]{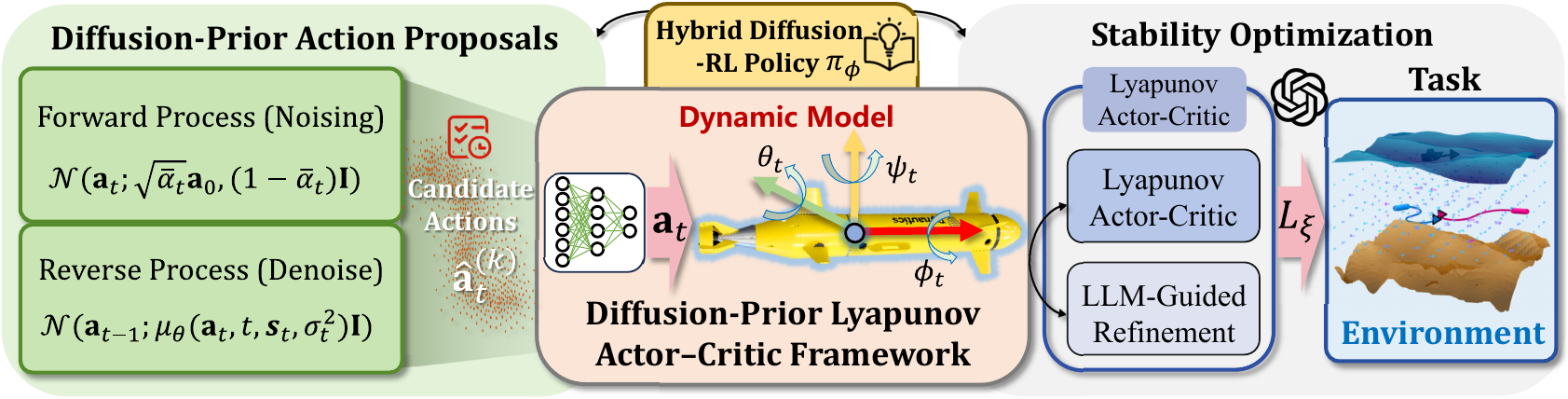}
        \caption{\small \textbf{Architecture of the proposed framework for AUV robust control}. This diffusion-prior Lyapunov actor-critic framework consists of three components: (A) Diffusion model for feasible action proposals; (B) Hybrid diffusion-RL policy, and (C) Lyapunov actor-critic with LLM-guided stability optimization.}
        \label{fig_2}
\end{figure*}

\section{METHODOLOGY}
In this section, we introduce the proposed framework in detail. As depicted in Fig. 2, this framework integrates diffusion models for generating long-horizon feasible actions, an RL backbone for policy optimization, and Lyapunov constraints with an LLM-driven outer loop for adaptive stability. Together, these modules form a generation–filtering–optimization pipeline that ensures performance, safety, and adaptability in dynamic requirements.

\subsection{Diffusion Model for Feasible Action Proposals}

Building upon recent advances in generative modeling, we formulate the problem of AUV control as a conditional denoising process that progressively refines noisy action sequences into feasible trajectories. Unlike RL, where exploration is injected via Gaussian noise, diffusion models explicitly learn the distribution of expert actions and can sample long-horizon, disturbance-resilient trajectories, enabling exploration that is both diverse and physically realizable \cite{18}.

\subsubsection{Forward and Reverse Processes}
To formalize this idea, we begin with the forward process, which gradually perturbs ground-truth actions with Gaussian noise. Given the action $\mathbf{a}_0 \in \mathbb{R}^d$ sampled from the dataset $\mathcal{D}$, noise is injected over $T$ steps, which can be calculated as follows:
\begin{equation}
q(\mathbf{a}_{1:T} \mid \mathbf{a}_0) = \prod_{t=1}^T q(\mathbf{a}_t \mid \mathbf{a}_{t-1}),
\end{equation}
with each transition defined as
\begin{equation}
q(\mathbf{a}_t \mid \mathbf{a}_{t-1}) \!=\! \mathcal{N}\!\left(\mathbf{a}_t;\sqrt{1-\beta_t}\mathbf{a}_{t-1}, \beta_t\mathbf{I}\right), \ t\!=\!1,\dots,T,
\end{equation}
where $\beta_t$ follows a linear schedule from $10^{-4}$ to $0.02$ over $T=1000$ steps, yielding a smooth transition from the data distribution to isotropic Gaussian noise. Here, $t$ denotes diffusion timesteps in the generative process, distinct from the environment timesteps of RL in Section~III-B. Equivalently, $\mathbf{a}_t$ can be sampled directly as
\begin{equation}
q(\mathbf{a}_t \mid \mathbf{a}_0) \!=\! \mathcal{N}\!\left(\mathbf{a}_t;\sqrt{\bar{\alpha}_t}\mathbf{a}_0,(1-\bar{\alpha}_t)\mathbf{I}\right), \ \bar{\alpha}_t\!=\!\prod_{i=1}^t(1-\beta_i).
\end{equation}

While the forward process destroys structure, the reverse process restores it by learning to denoise trajectories back to the data distribution. This is parameterized by a neural network $\epsilon_\theta(\cdot)$ conditioned on the AUV state. In practice, we implement $\epsilon_\theta$ as a U-Net architecture that incorporates both spatial and temporal information \cite{U-Net}. To achieve this, the input is constructed by concatenating an MLP-encoded state representation with a sinusoidal time embedding:
\begin{equation}
\epsilon_\theta(\mathbf{a}_t, t, \mathbf{s}_t) = \text{U-Net}\!\left(\text{Concat}\big[\text{MLP}(\mathbf{s}_t), \text{TimeEmb}(t)\big]\right).
\end{equation}

Here, the state $\mathbf{s}_t$ integrates current observations with historical state–action context:
\begin{equation}
\mathbf{s}_t = [\mathbf{o}_t, \text{Flatten}(\mathbf{H}_{t-L:t-1}), \text{Flatten}(\mathbf{A}_{t-L:t-1})],
\end{equation}
where $L=10$ balances temporal coverage with computational efficiency, and the flatten operation preserves sequential correlations while reducing dimensionality. Meanwhile, the time embedding encodes the diffusion step as
\begin{equation}
\text{TimeEmb}(t) = \text{MLP}(\sin(10^4 t / N)), \quad N = 1000,
\end{equation}
providing smooth, periodic representations that help the model distinguish noise levels across denoising stages. 

With these inputs, the U-Net learns to parameterize the reverse denoising distribution, which can be expressed as
\begin{equation}
p_\theta(\mathbf{a}_{0:T}) = p(\mathbf{a}_T) \prod_{t=1}^T p_\theta(\mathbf{a}_{t-1} \mid \mathbf{a}_t, \mathbf{s}_t),
\end{equation}
where the\! process starts from $p(\mathbf{a}_T)\!\!=\!\!\mathcal{N}\!(0,\!\mathbf{I})$,\! treating corrupted actions as noise, with each reverse step parameterized as
\begin{equation}
p_\theta(\mathbf{a}_{t-1} \mid \mathbf{a}_t, \mathbf{s}_t) = \mathcal{N}\!\left(\mathbf{a}_{t-1}; \mu_\theta(\mathbf{a}_t,t,\mathbf{s}_t), \sigma_t^2 \mathbf{I}\right).
\end{equation}

Moreover, we train the reverse model with a denoising score matching loss that aligns predicted noise with true Gaussian noise:
\begin{equation}
\mathcal{L}_{\text{diff}}(\theta) = 
\mathbb{E}_{\mathbf{a}_0,\epsilon,t}\!\Big[ 
\| \epsilon - \epsilon_\theta(\sqrt{\bar{\alpha}_t}\mathbf{a}_0 + \sqrt{1-\bar{\alpha}_t}\epsilon, t, \mathbf{s}_t) \|^2 
\Big],
\end{equation}
where $\epsilon \sim \mathcal{N}(0,\mathbf{I})$ is Gaussian noise added at step $t$.

\subsubsection{Training and Inference}

With these components in place, training proceeds by sampling a random timestep $t$, perturbing ground-truth actions with Gaussian noise, and minimizing $\mathcal{L}_{\text{diff}}(\theta)$. At inference time, the model begins with Gaussian noise $\mathbf{a}_T \sim \mathcal{N}(0,\mathbf{I})$ and iteratively applies the reverse process:
\begin{equation}
\mathbf{a}_{t-1} \!\!=\!\! \frac{1}{\sqrt{1\!-\!\beta_t}}\!\Big(\mathbf{a}_t \!-\! \frac{\beta_t}{\sqrt{1\!-\!\bar{\alpha}_t}} \epsilon_\theta(\mathbf{a}_t,t,\mathbf{s}_t)\Big) \!+\! \sigma_t \mathbf{z}, \ \mathbf{z}\!\sim\!\mathcal{N}(0,\mathbf{I}).
\end{equation}
After $T$ steps, the final denoised action sequence $\mathbf{a}_0$ is obtained as a candidate action proposal.

In summary, the diffusion model combines forward perturbation, reverse denoising, state conditioning, and temporal encoding to generate feasible action trajectories. Compared with naive noise injection, this approach produces multimodal, smooth, and disturbance-resilient exploration proposals.

\subsection{Hybrid Diffusion-Reinforcement Learning Policy}
While RL offers strong reward-driven optimization, its limited ability in long-horizon planning often yields myopic behavior under sparse rewards. Diffusion models, in turn, generate temporally consistent action sequences with long-range dependencies but remain agnostic to task rewards. We unify these perspectives in a single framework, where diffusion shapes the trajectory space toward long-horizon coherence, while RL aligns it with task-specific rewards. Finally, the resulting policy $\pi_\phi$ achieves both temporal foresight and reward refinement, supported by a low-level S\mbox{-}Surface controller for control signal execution.

At the foundation, we model AUV control as an infinite-horizon discounted Markov Decision Process (MDP) 
\(\mathcal{M}=(\mathcal{S},\mathcal{A},\mathcal{P},r,\gamma)\), 
where \(\mathcal{S}\) is the state space, \(\mathcal{A}\) the action space, 
\(\mathcal{P}(\mathbf{s'}\!\mid \mathbf{s},\mathbf{a})\) the transition kernel, 
\(r:\mathcal{S}\times\mathcal{A}\to\mathbb{R}\) the immediate reward, 
and \(\gamma\in[0,1)\) the discount factor. 
A stochastic policy \(\pi_\phi(\mathbf{a}\mid \mathbf{s})\) parameterized by \(\phi\) aims to maximize the expected return, which can be expressed as follows \cite{TMC}:
\begin{equation}
J(\pi_\phi)=\mathbb{E}_{\vartheta\sim(\pi_\phi,\mathcal{P})}\!\left[\sum_{t=0}^{\infty}\gamma^{t}\,r(\mathbf{s}_t,\mathbf{a}_t)\right],
\end{equation}
where \(\mathbf{s}_t\in\mathcal{S}\) and \(\mathbf{a}_t\in\mathcal{A}\) denote the state and action at time \(t\), 
and the expectation is taken over trajectories \(\vartheta=(\mathbf{s}_0,\mathbf{a}_0,\mathbf{s}_1,\dots)\) induced by \(\pi_\phi\) and \(\mathcal{P}\).

Then, the optimization follows the policy gradient theorem:
\begin{equation}
\nabla_\phi J(\pi_\phi)=\mathbb{E}_{\mathbf{s}\sim\rho^\pi,\mathbf{a}\sim\pi_\phi}\!\big[\nabla_\phi\log \pi_\phi(\mathbf{a}\mid \mathbf{s})\,Q^\pi(\mathbf{s},\mathbf{a})\big],
\end{equation}
where $\rho^\pi$ denotes the stationary state distribution and $Q^\pi(\mathbf{s},\mathbf{a})$ is the action-value function under policy $\pi$. To mitigate the high variance of direct gradients, we adopt an actor–critic scheme: the actor outputs actions, while the critic estimates returns by minimizing the TD error
\begin{equation}
\mathcal{L}_Q(\psi)=\mathbb{E}_{(\mathbf{s},\mathbf{a},r,\mathbf{s'})}\!\Big[(Q_\psi(\mathbf{s},\mathbf{a})-y)^2\Big], \ y=r+\gamma V^\pi(\mathbf{s'}).
\end{equation}
where $V^\pi(\mathbf{s'})$ denotes the state-value function, i.e., the expected return from state $\mathbf{s'}$ under policy $\pi$.

Based on the above foundations, we further integrate a diffusion model that generates $K$ diverse candidate actions at each timestep, conditioned on the current state:
\begin{equation}
\hat{\mathbf{a}}_t^{(k)} \sim p_\theta(\mathbf{a}_0\mid \mathbf{s}_t), \quad k=1,\dots,K,
\end{equation}
where $p_\theta(\mathbf{a}_0 \!\!\mid \!\!\mathbf{s}_t)$ represents the marginal distribution of denoised actions obtained via reverse diffusion. Furthermore, the critic evaluates these candidates and selects the one with the highest expected return:
\begin{equation}
\pi_\phi(\mathbf{s}_t)=\mathop{\arg\max}\limits_{k\in\{1,\dots,K\}}Q_\psi(\mathbf{s}_t,\hat{\mathbf{a}}_t^{(k)}).
\end{equation}
This hybrid mechanism balances diffusion’s ability to propose long-horizon, disturbance-resilient actions with RL’s capacity to optimize for long-term cumulative reward.

To close the loop between high-level policy and physical execution, we integrate the S\mbox{-}Surface controller with a six-degree-of-freedom (6-DoF) AUV dynamic model \cite{Oceans}. The S\mbox{-}Surface control law is defined as
\begin{equation}
u_t=\frac{2}{1+\exp(-\zeta_1 e-\zeta_2 \dot{e})}-1+\delta u,
\end{equation}
where $e$ and $\dot{e}$ denote the tracking error and its derivative, $\zeta_1,\zeta_2$ are nonlinear shaping gains, and $\delta u$ stands for a compensation term for real-time disturbance rejection. This formulation ensures smooth outputs and adaptive gain tuning.

To faithfully capture how these control signals interact with the vehicle’s physical dynamics, 
we adopt Fossen’s 6-DoF model~\cite{Oceans2}. 
The equations of motion are
\begin{equation}
\boldsymbol{M}_i\dot{\boldsymbol{v}}_i+\boldsymbol{C}_i(\boldsymbol{v}_i)\boldsymbol{v}_i+\boldsymbol{D}_i(\boldsymbol{v}_i)\boldsymbol{v}_i+\boldsymbol{G}_i(\boldsymbol{\eta}_i)=\boldsymbol{\tau}_i,
\end{equation}
with kinematics
\begin{equation}
\dot{\boldsymbol{\eta}}_i=\boldsymbol{J}(\boldsymbol{\eta}_i)\boldsymbol{v}_i,
\end{equation}
where $\boldsymbol{v}_i$ denotes the body-fixed velocity vector, while $\boldsymbol{\eta}_i$ is the position–orientation vector in the inertial frame. Besides, $\boldsymbol{M}_i$ represents the mass matrix, $\boldsymbol{C}_i$ represents the Coriolis terms, $\boldsymbol{D}_i$ denotes damping, $\boldsymbol{G}_i$ stands for restoring forces, and $\boldsymbol{J}(\cdot)$ is the transformation matrix. The control input $\boldsymbol{\tau}_i$ is generated by mapping the S\mbox{-}Surface outputs $u_t$ into generalized forces and moments through the actuator model.

For discrete-time training, the model is discretized as
\begin{subequations}
\begin{align}
\boldsymbol{\eta}_{t+1} &= \boldsymbol{\eta}_t+\Delta t\cdot\boldsymbol{J}(\boldsymbol{\eta}_t)\boldsymbol{v}_t, \\
\boldsymbol{v}_{t+1} &= \boldsymbol{v}_t+\Delta t\,\boldsymbol{M}_i^{-1}F(\boldsymbol{\eta}_t,\boldsymbol{v}_t).
\end{align}
\end{subequations}
with residual force
\begin{equation}
F(\boldsymbol{\eta}_t,\boldsymbol{v}_t)=\boldsymbol{\tau}_t-\boldsymbol{C}(\boldsymbol{v}_t)\boldsymbol{v}_t-\boldsymbol{D}(\boldsymbol{v}_t)\boldsymbol{v}_t-\boldsymbol{G}(\boldsymbol{\eta}_t).
\end{equation}

Altogether, the hybrid diffusion-RL framework unifies diffusion-enhanced exploration, actor-critic optimization, and S\mbox{-}Surface-based low-level control on a realistic 6-DoF AUV model. This design not only ensures diverse and robust action proposals but also grounds them in long-term stability and physical feasibility, enabling adaptive and reliable AUV operation in challenging ocean environments.

\subsection{Lyapunov Actor-Critic with LLM-Guided Stability Optimization}
While diffusion-augmented RL enhances exploration and improves long-horizon planning, it does not inherently guarantee the stability of the resulting policies. To bridge this gap, we adopt the Lyapunov Actor-Critic (LAC) framework, which incorporates explicit stability constraints into the learning process \cite{LAC}. Yet, the effectiveness of LAC critically depends on the choice of Lyapunov functions, which are often problem-specific and brittle when transferred across task requirements. To overcome this limitation, we further introduce an LLM-driven outer loop that automatically selects and refines Lyapunov functions based on task semantics and training feedback. In this way, diffusion, LAC, and LLM modules are integrated into a unified framework, where exploration, stability, and adaptability are simultaneously enforced. 

\subsubsection{Lyapunov Actor--Critic Formulation}
The key principle of the LAC framework in this study is to encode stability through a cost-based Lyapunov function $L_\xi(\mathbf{s},\mathbf{a})$, parameterized by $\xi$. 
Concretely, we approximate $L_\xi(\mathbf{s},\mathbf{a})$ with a neural critic network and enforce it to predict the discounted cumulative reward along the system trajectories.
Given an instantaneous reward $r(\mathbf{s},\mathbf{a})$, the Lyapunov critic is trained to satisfy a Bellman-type recursion:
\begin{equation}
    L_\xi(\mathbf{s},\mathbf{a}) \approx -r(\mathbf{s},\mathbf{a}) + \gamma \, \mathbb{E}_{\mathbf{s}'}\big[L_\xi\big(\mathbf{s}',\pi(\mathbf{s}')\big)\big],
    \label{eq:our_lyap_bellman}
\end{equation}
where $\mathbf{s}'$ is the successor state after applying the action $\mathbf{a}$, and $\pi$ is the current policy.
In implementation, the critic network outputs a scalar that is passed through a softplus activation to guarantee positivity, i.e.,
\begin{equation}
    L_\xi(\mathbf{s},\mathbf{a}) = \mathrm{softplus}\big(f_\xi(\mathbf{s},\mathbf{a})\big) > 0,
    \label{eq:our_lyap_softplus}
\end{equation}
and is optimized by minimizing the mean squared temporal-difference error between $L_\xi(\mathbf{s},\mathbf{a})$ and the target
\begin{equation}
    \hat{L}(\mathbf{s},\mathbf{a}) = -r(\mathbf{s},\mathbf{a}) + \gamma (1-d)\,
    L_{\xi^-}\big(\mathbf{s}',\pi(\mathbf{s}')\big),
\end{equation}
where $\xi^-$ denotes the parameters of the target Lyapunov critic, and $d$ is the terminal flag.

On top of this Lyapunov critic, the actor is updated by explicitly imposing a Lyapunov stability condition.
Given a batch of transitions $(\mathbf{s},\mathbf{a},r,\mathbf{s}')$, we evaluate the Lyapunov value of the current action $\mathbf{a}$ and that of the new action proposed by the actor $\pi$:
\begin{equation}
    L_{\text{old}} = L_\xi(\mathbf{s},\mathbf{a}), \qquad
    L_{\text{new}} = L_\xi\big(\mathbf{s},\pi(\mathbf{s})\big).
\end{equation}
We then define a stability condition that combines the Lyapunov difference and the instantaneous reward:
\begin{equation}
    \Delta L(\mathbf{s}) 
    = L_{\text{new}} - L_{\text{old}} - \alpha \, r(\mathbf{s},\mathbf{a}),
    \label{eq:our_stability_condition}
\end{equation}
where $\alpha > 0$ is a tunable stability coefficient.
Intuitively, Eq. (25) penalizes policies that both increase the predicted Lyapunov value and incur a high immediate cost.

The actor update is formulated as a constrained optimization problem:
\begin{align}
    \min_{\pi} \quad
        & \mathbb{E}_{\mathbf{s}}\big[ \Delta L(\mathbf{s}) \big] 
        \;+\; \beta \, \mathbb{E}_{s}\big[ \log \pi(\mathbf{a}|\mathbf{s}) + 1 \big],
        \label{eq:our_actor_obj}
        \\
    \text{s.t.} \quad
        & \mathbb{E}_{\mathbf{s}}\big[ \Delta L(\mathbf{s}) \big] \le 0,
        \label{eq:our_actor_constraint}
\end{align}

To solve Eqs. (26)--(27), we introduce dual variables $\lambda \ge 0$ and $\beta \ge 0$ and optimize the following Lagrangian:
\begin{equation}
\mathcal{L}(\pi,\xi,\lambda,\beta)
=
\mathbb{E}_{\mathbf{s}}\!\left[
    \lambda \,\Delta L(\mathbf{s})
    + \beta \,\big(\log \pi(\mathbf{a}|\mathbf{s}) + 1\big)
\right],
\label{eq:our_lac_lagrangian}
\end{equation}
where $\lambda = \exp(\log\lambda)$ and $\beta = \exp(\log\beta)$ are updated by gradient ascent on their corresponding dual objectives, which encourage the policy to reduce the Lyapunov value and cost while maintaining sufficient exploration.

\subsubsection{LLM-Guided Lyapunov Function Generation}

Despite its theoretical guarantees, the success of LAC hinges on an appropriate choice of $L_\xi(\mathbf{s})$. Different task requirements often prioritize various stability aspects, making fixed functions insufficient \cite{Lyapunov}. To address this, we leverage an LLM to guide the generation of Lyapunov functions. 

We begin with a library of candidate forms:
\begin{equation}
\mathcal{F} = \{ f_1(\mathbf{s}), f_2(\mathbf{s}), \dots, f_m(\mathbf{s}) \},
\end{equation}

Instead of fixing one function a priori, the LLM acts as a semantic selector:
\begin{equation}
L_\phi(\mathbf{s}) \approx f_j(\mathbf{s}), \quad f_j \in \mathcal{F}, \quad j = \text{LLM}(\mathcal{T}),
\end{equation}
where $\mathcal{T}$ denotes the task description. This ensures that the Lyapunov function is aligned with mission requirements at the onset of training.

Moreover, the training feedback, such as frequent constraint violations or episodes of instability, can prompt the LLM to switch functions within $\mathcal{F}$. For example, if velocity oscillations dominate, the LLM may recommend moving from a position-based Lyapunov function to a velocity-oriented one, thereby adapting stability enforcement to the current dynamics.

Overall, in this diffusion-prior LAC framework, the diffusion model provides diverse candidate actions; LAC filters them through stability constraints, and the LLM adaptively maintains the semantic alignment of the Lyapunov function with mission objectives. This integration achieves two goals simultaneously: (1) \textbf{Improvement of stability}, ensuring reliable convergence of policy updates while penalizing actions that increase Lyapunov value;  
(2) \textbf{Task-specific adaptability}, as the LLM flexibly adjusts Lyapunov functions to dynamic requirements. Finally, the pseudo-code of the proposed framework is shown in Algorithm 1.

\begin{algorithm}[!t]
\caption{Diffusion-Prior LAC Framework}
\begin{algorithmic}[1]
\STATE \textbf{Initialize:} Diffusion model $p_\theta(\mathbf{a}_0|\mathbf{s})$, actor $\pi_\phi$, critic $Q_\psi$, Lyapunov critic $L_\xi$, LLM module, and replay buffer $\mathcal{R}$
\FOR{each training episode}
  \STATE Sample initial state $\mathbf{s}_0$
  \FOR{each timestep $t$}
    \STATE \textbf{Diffusion generation:} 
    Sample $K$ candidate actions $\{\hat{\mathbf{a}}_t^{(k)}\}$ from $p_\theta(\mathbf{a}_0|\mathbf{s}_t)$
    \STATE \textbf{RL filtering:} 
    Select $\mathbf{a}_t = \arg\max_k Q_\psi(\mathbf{s}_t,\hat{\mathbf{a}}_t^{(k)})$
    \STATE Execute $\mathbf{a}_t$, observe $(r_t, \mathbf{s}_{t+1})$, and store transition in replay buffer $\mathcal{R}$
    \IF{update step}
      \STATE Update critic $Q_\psi$ and actor $\pi_\phi$ using sampled batches from replay buffer $\mathcal{R}$
      \STATE \textbf{Stability enforcement:} 
      Constrain policy update with Lyapunov critic $L_\xi$
      \STATE \textbf{LLM adaptation:} 
      If instability or constraint violation detected, query LLM to refine $L_\xi$ based on task semantics
      \STATE Soft update target networks $Q_{\psi'}$, $L_{\xi'}$
    \ENDIF
  \ENDFOR
\ENDFOR
\end{algorithmic}
\end{algorithm}

\section{EXPERIMENTS}
In this section, we perform extensive simulations to validate the effectiveness of the proposed framework, followed by the experimental results and analysis.
\subsection{Task Description and Settings} 
To thoroughly evaluate the proposed framework, we employ the REMUS-100 AUV (1.6 m, 31.9 kg) as the experimental AUV model and construct a realistic 3D underwater data-collection environment, given the lack of standardized benchmarks for this task. As illustrated in Fig. 3, an Autonomous Surface Vehicle (ASV) operates as a mobile communication relay and positioning anchor for the underwater agents, enabling stable acoustic links and enhancing localization accuracy. Multiple AUVs equipped with our proposed framework collaboratively navigate in the scenario to collect data from spatially distributed Sensor Nodes (SNs) within an Internet of Underwater Things (IoUT) environment. 

\begin{table}[!t]
  \centering
  \caption{\small Key Parameters and Hyperparameters Configuration} 
  \label{tab:hyperparameters}
  
  \begin{tabular}{lc}
    \toprule
    \textbf{Parameters} & \textbf{Values} \\
    \midrule
    Diffusion steps ($T$) & 1000 \\
    Denoising steps & 50 \\
    Noise schedule ($\beta_t$) & Linear $1\times10^{-4}\rightarrow0.02$ \\
    History length ($L$) & 10 \\
    Candidate actions ($K$) & 5 \\
    U-Net hidden dimension & 256 \\
    Training batch size (RL/Diffusion) & 64 / 32 \\
    Hidden layer size (RL/Diffusion) & 128 / 256 \\
    Learning rate (RL/Diffusion) & $1\times10^{-3}$ / $1\times10^{-4}$ \\
    Discount factor ($\gamma$) & 0.97 \\
            AUV maximum speed ($v_{\text{max}}$, $\omega_{\text{max}}$) & 2.3m/s, 0.26rad/s\\
        Propeller maximum revolution & 1525rpm \\
        Water density ($\rho$) & 1026$\text{kg/m}^{\text{3}}$ \\
        Control frequency & 20Hz \\
        Controller parameters for yaw ($\zeta_1,\zeta_2$) & 2, 1 \\
        Controller\! parameters for\! depth ($\zeta_1,\zeta_2$) & 1, 1 \\
    LLM model & GPT-4o \\
    LLM parameters & temperature=0.5, Top P=1 \\
    \bottomrule
  \end{tabular}
\end{table}

\begin{figure}[!t]
        \centering
        \includegraphics[width=0.99\linewidth]{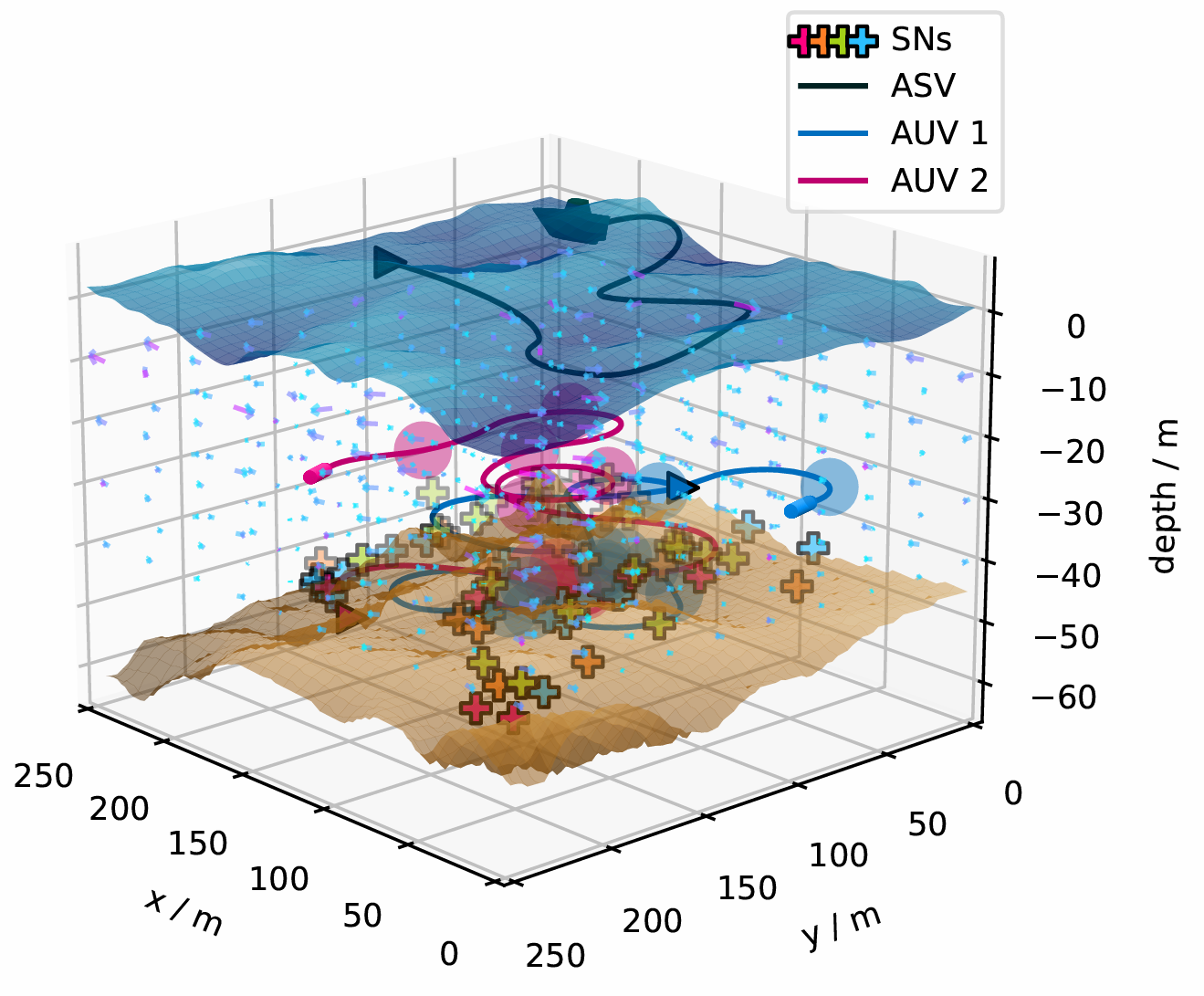}
        \caption{\small Visualizations of the scenario adopted in this study, where the ASV serves as a mobile communication relay and positioning anchor for the underwater AUVs, and multiple AUVs equipped with the proposed framework collaboratively navigate the environment to perform the data collection task.}
        \label{fig_3a}
\end{figure}

The scene features a dynamic sea surface, heterogeneous bathymetry, and SNs scattered throughout the 3D volume, while the AUVs follow coordinated trajectories to fulfill the mission. The system jointly optimizes several objectives: maximizing the number of Serviced Sensor Nodes (SSN) and the overall Sum Data Rate (SDR), while simultaneously minimizing collision risks and total Energy Consumption (EC). The key parameter settings are summarized in TABLE I, with additional simulation configurations included in the previous work \cite{20}.

Finally, simulations were conducted on a Ryzen~9 5950X CPU and RTX~3060 GPU using Python~3.12. 
The system completed 300 episodes of training in around 8 hours, reflecting reasonable computational efficiency.

\subsection{Results and Analysis} 
As the first stage of our experiments, we begin by training the diffusion model using offline state-action demonstrations. The model takes short sequences of states and corresponding action segments, samples a random diffusion timestep~$t$, perturbs the actions using the cosine-based diffusion schedule $\sqrt{\bar{\alpha}_t}$ and $\sqrt{1-\bar{\alpha}_t}$, and is trained via an MSE objective to predict the injected Gaussian noise while conditioning on the normalized timestep. This procedure enables the model to learn time-aware denoising dynamics and capture the underlying multi-step action distribution. The training curve in Fig.~4 exhibits a rapid loss reduction during the early phase, followed by a smooth convergence over $10^5$ epochs, with the inset showing the loss stabilizing around $3\times10^{-2}$ in the final stage. These results indicate that the diffusion model successfully learns consistent temporal correlations in the action sequences, allowing it to later generate smooth, robust, and dynamically feasible control commands for underwater AUV control within our framework.

\begin{figure}[!t]
        \centering
        \includegraphics[width=0.99\linewidth]{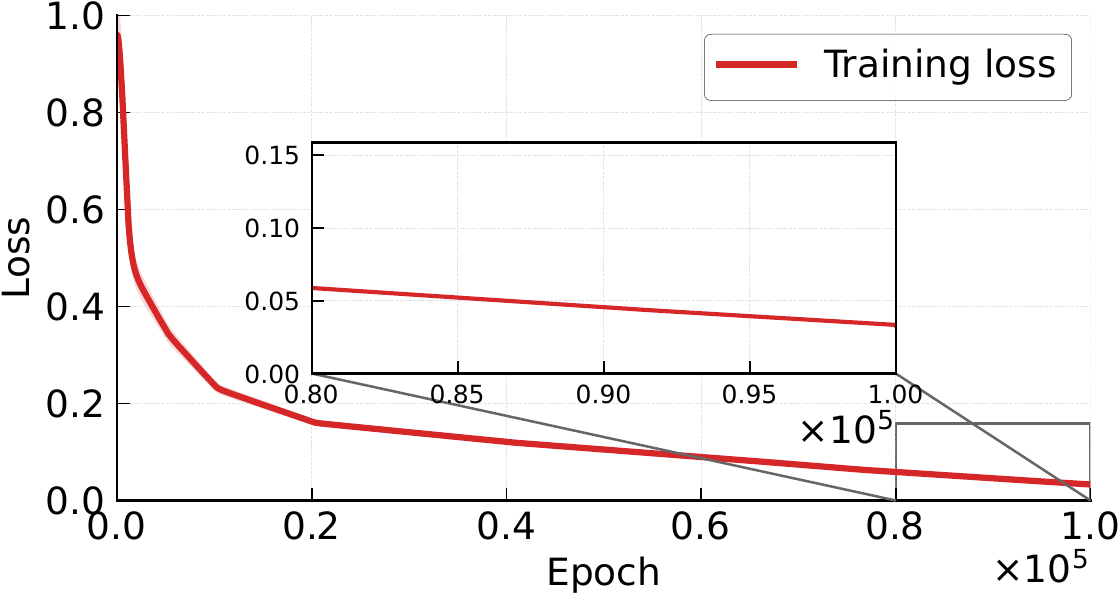}
        \caption{\small Training loss curve of the diffusion model, showing rapid initial decay and smooth long-term convergence, with an inset highlighting the final stabilization phase around $3\times10^{-2}$.}
        \label{fig_3}
\end{figure}

\begin{figure*}[!t]
        \centering
        \includegraphics[width=0.99\linewidth]{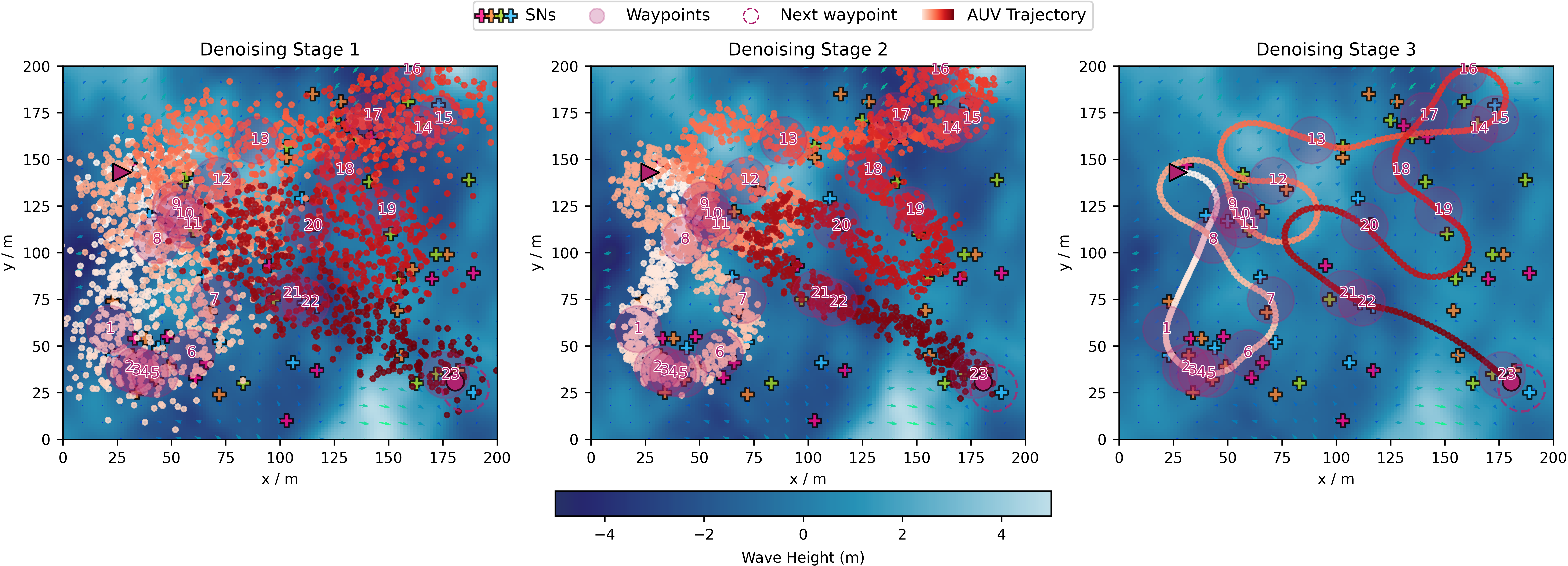}
        \caption{\small Visualization of five candidate actions across three denoising stages under the proposed framework. Trajectories evolve from scattered exploration to smooth, task-aligned paths, demonstrating the ability of diffusion model to generate diverse yet optimized plans.}
        \label{fig_10}
\end{figure*}

\begin{figure*}[!t]
        \centering
        \includegraphics[width=0.99\linewidth]{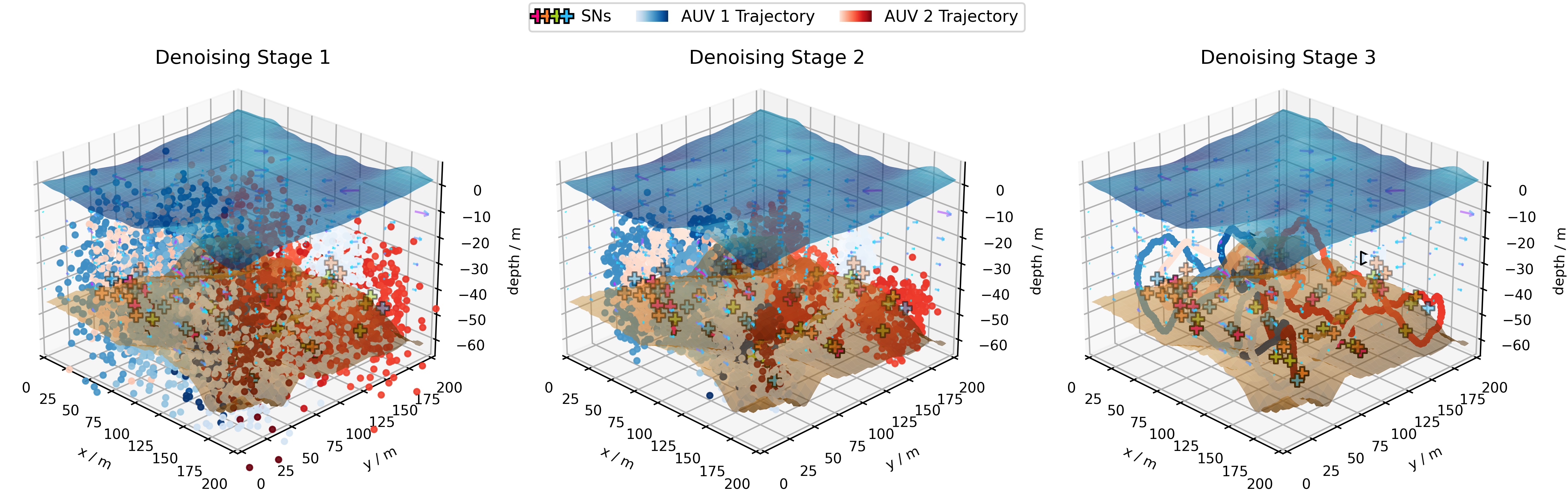}
        \caption{\small 3D trajectory evolution across denoising stages under the proposed framework. Candidates transition from scattered to coordinated paths, enabling robust multi-AUV planning and robust control.}
        \label{fig_11}
\end{figure*}

After establishing a well-trained diffusion model, we next illustrate how it operates within our proposed framework. As shown in Fig.~5 and Fig.~6, the mechanism is visualized by executing five candidate actions sampled at different stages of the reverse diffusion process. Each subplot corresponds to a specific denoising step, in which the five candidates are independently rolled out as colored trajectories. In the early stage of denoising, the trajectories appear highly scattered, both in the 2D projections of a single AUV in Fig.~5 and in the 3D spatial evolution of two AUVs in Fig.~6, reflecting the exploratory behavior induced by strong noise. As the reverse process progresses, these initially dispersed trajectories gradually become more structured, beginning to align with task-relevant waypoints, ocean topography, and sensor node distributions. By the final stage, the candidates converge into smooth, goal-directed, and dynamically feasible paths that remain well separated across the two AUVs, thereby supporting coordinated behaviors. This evolution—from noise-driven diversity to progressively refined and task-aware action quality—highlights the diffusion model’s ability to generate diverse yet coherent multi-step trajectories, enabling multi-AUV cooperation and providing planning capabilities that conventional RL struggles to achieve in sparse-reward and highly dynamic underwater environments.

To evaluate the effectiveness of the proposed framework (hereafter referred to as the Diffusion+RL framework), we compare its performance against a standard RL baseline under ideal conditions, Extreme Sea conditions (ES), and Very Extreme Sea conditions (VES) for the underwater data collection task. As shown in Fig.~7, all Diffusion+RL variants, including the LAC-augmented version Diffusion+RL (LAC), exhibit substantially faster convergence than standard RL, reaching stable performance within roughly 300 episodes and demonstrating markedly improved sample efficiency. In terms of mission-related metrics, Diffusion+RL and Diffusion+RL (LAC) achieve higher SDR and consistently serve more SSN, with the Diffusion+RL (LAC) providing an additional improvement under ES and VES disturbances. EC remains comparable across all diffusion-based methods, indicating that the gains in coverage and throughput do not incur additional energy costs. Finally, the episode return curves show that Diffusion+RL and Diffusion+RL (LAC) reach higher long-term returns than the baseline, especially in challenging sea states, reflecting more robust and well-structured control behavior. Overall, these results demonstrate that diffusion-driven action generation, which is further strengthened by LAC, accelerates learning and yields more effective and resilient control policies in complex and highly dynamic underwater environments.

\begin{figure*}[!t]
        \centering
        \includegraphics[width=0.99\linewidth]{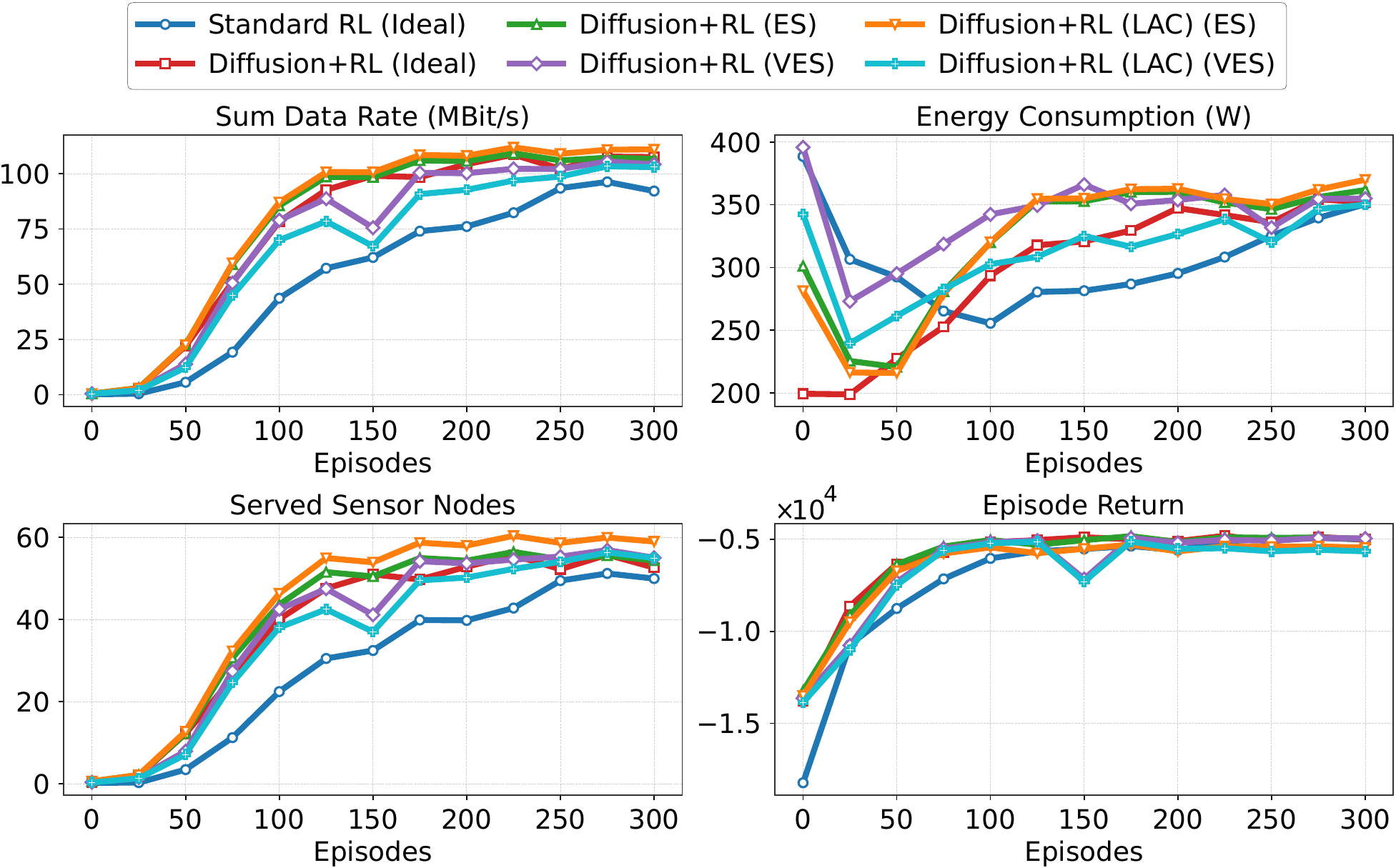}
        \caption{\small Performance comparison between standard RL, diffusion+RL and diffusion+RL (LAC) across underwater data collection task metrics under ideal, ES and VES conditions, respectively.}
        \label{fig_4}
\end{figure*}

Furthermore, we investigate how different low-level controllers influence the overall performance of the proposed framework, considering three commonly used controllers---S-Surface, PID, and SMC---evaluated under both ES and VES conditions. As summarized in TABLE~II, clear performance differences emerge across controllers and sea conditions. Within the standard Diffusion+RL setting, the S-Surface controller provides the strongest results, achieving an SDR of $105.1 \pm 14.7$~MBit/s under ES and $101.0 \pm 7.5$~MBit/s under VES, while also serving the highest number of SNs (up to $55.4 \pm 9.5$). PID exhibits moderate yet stable performance, delivering SDR values around $88.6 \pm 7.2$ (ES) and $86.1 \pm 7.6$~MBit/s (VES), though with slightly elevated energy consumption. In contrast, SMC performs poorly in challenging conditions, showing severe degradation under VES, with SDR dropping to $12.8 \pm 5.2$~MBit/s and SSN to only $6.6 \pm 2.7$, indicating limited robustness to compounded environmental disturbances. When incorporating LAC into Diffusion+RL, the overall performance further improves. Notably, Diffusion+RL (LAC) with the S-Surface controller yields the highest SDR among all configurations---$110.6 \pm 13.5$~MBit/s (ES) and $102.3 \pm 8.4$~MBit/s (VES)---and consistently provides the greatest SN coverage (up to $57.1 \pm 9.3$). PID and SMC also benefit from LAC, showing higher robustness and reduced energy consumption compared to their non-LAC counterparts. These results collectively demonstrate that (i) the S-Surface controller synergizes best with diffusion-driven high-level planning and (ii) the inclusion of LAC enhances stability and resilience under extreme and highly dynamic sea conditions, enabling more reliable task execution in real-world underwater environments.

Building upon these results, we further analyze how different low-level controllers affect tracking accuracy within the diffusion-based frameworks. Table~III compares yaw and depth tracking errors across S-Surface, PID, and SMC controllers under four settings: Diffusion+RL, Diffusion, Diffusion+RL (LAC), and the RL baseline. A consistent pattern emerges across all configurations: the S-Surface controller achieves the lowest tracking errors in both yaw and depth. For example, under Diffusion+RL, S-Surface attains a yaw error of $\mu = 0.12$, $\sigma = 0.06$ and a depth error of $\mu = 1.59$, $\sigma = 1.03$, outperforming both PID and SMC. This advantage becomes even more pronounced when LAC is included, where S-Surface achieves the best results in the entire table, with the lowest yaw error ($\mu = 0.11$, $\sigma = 0.05$) and the most stable depth tracking ($\mu = 1.52$, $\sigma = 0.94$). In contrast, PID exhibits larger steady-state deviations and higher variance across all frameworks, while SMC suffers from higher depth errors and inconsistent yaw tracking, especially under Diffusion and baseline conditions. These findings indicate that S-Surface provides stronger disturbance rejection and nonlinear stabilization properties, and when combined with diffusion-guided action generation, it enables the most precise and consistent trajectory tracking in dynamic underwater environments.

\begin{table*}
  \centering
  \caption{\small Performance Comparison of Two Frameworks Using Different Low-Level Controllers\\under ES and VES Conditions} 
  \begin{tabular}{ccc ccc}
      \toprule
      \textbf{Framework} &
      \textbf{Controller} &
      \textbf{Condition} &
      \textbf{SDR} (MBit/s) $\uparrow$ & 
      \textbf{EC} (W) $\downarrow$ & 
      \textbf{SSN} $\uparrow$ \\
      \midrule

      \multirow{6}{*}{\raisebox{-0.7em}{Diffusion + RL}}
        & \multirow{2}{*}{S-Surface}
            & ES  & 105.1 $\pm$ 14.7 & 362.5 $\pm$ 22.8 & 55.4 $\pm$ 9.5 \\
        &     & VES & 101.0 $\pm$ 7.5  & 352.2 $\pm$ 6.5  & 54.4 $\pm$ 6.5 \\ 
      \cmidrule(lr){2-6}

        & \multirow{2}{*}{PID} 
            & ES  & 88.6 $\pm$ 7.2  & 257.1 $\pm$ 17.6 & 44.5 $\pm$ 4.3 \\
        &     & VES & 86.1 $\pm$ 7.6  & 246.7 $\pm$ 20.5 & 43.6 $\pm$ 4.6 \\
      \cmidrule(lr){2-6}

        & \multirow{2}{*}{SMC} 
            & ES  & 41.9 $\pm$ 7.0  & 225.7 $\pm$ 13.5 & 22.4 $\pm$ 3.1 \\
        &     & VES & 12.8 $\pm$ 5.2  & 173.0 $\pm$ 11.8 & 6.6 $\pm$ 2.7 \\
      \cmidrule(lr){1-6}
      
        \multirow{6}{*}{\raisebox{-0.7em}{Diffusion + RL (LAC)}}
        & \multirow{2}{*}{S-Surface}
            & ES  & \textbf{110.6} $\pm$ \textbf{13.5} & \textbf{371.9} $\pm$ \textbf{23.5} & \textbf{57.1} $\pm$ \textbf{9.3} \\
        &     & VES & \textbf{102.3} $\pm$ \textbf{8.4}  & \textbf{349.0} $\pm$ \textbf{6.3}  & \textbf{54.8} $\pm$ \textbf{5.9} \\ 
      \cmidrule(lr){2-6}

        & \multirow{2}{*}{PID} 
            & ES  & 93.3 $\pm$ 6.1  & 265.2 $\pm$ 16.2 & 45.8 $\pm$ 4.4 \\
        &     & VES & 89.9 $\pm$ 7.2  & 250.4 $\pm$ 18.7 & 44.0 $\pm$ 4.2 \\
      \cmidrule(lr){2-6}

        & \multirow{2}{*}{SMC} 
            & ES  & 50.2 $\pm$ 8.1  & 240.5 $\pm$ 17.8 & 25.4 $\pm$ 4.0 \\
        &     & VES & 23.7 $\pm$ 6.6  & 201.8 $\pm$ 12.9 & 13.4 $\pm$ 4.5 \\
      \bottomrule 
  \end{tabular}
\end{table*}

\begin{table*}[t]
\centering
\caption{\small Yaw and Depth Tracking Errors of Four Frameworks Using Different Low-level Controllers}
\begin{tabular}{l lcccc}
\toprule
\multirow{2}{*}{\textbf{Framework}} &
\multirow{2}{*}{\textbf{Controller}} &
\multicolumn{2}{c}{\textbf{Yaw Error (rad)}} &
\multicolumn{2}{c}{\textbf{Depth Error (m)}} \\
\cmidrule(r){3-4} \cmidrule(r){5-6}
 & & \textbf{Mean ($\mu$)} & \textbf{Standard Deviation ($\sigma$)}  & \textbf{Mean ($\mu$)} & \textbf{Standard Deviation ($\sigma$)}  \\
\midrule
\multirow{3}{*}{Diffusion + RL} 
    & SMC        & 0.25 & 0.13 & 2.05 & 1.49 \\
    & PID        & 0.51 & 0.37 & 1.95 & 1.31 \\
    & S-Surface  & 0.12 & 0.06 & 1.59 & 1.03 \\
\midrule
\multirow{3}{*}{Diffusion} 
    & SMC        & 0.28 & 0.14 & 2.17 & 1.48 \\
    & PID        & 0.57 & 0.40 & 2.08 & 1.39 \\
    & S-Surface  & 0.13 & 0.05 & 1.67 & 1.15 \\
\midrule
\multirow{3}{*}{Diffusion + RL (LAC)} 
    & SMC        & 0.23 & 0.10 & 1.89 & 1.42 \\
    & PID        & 0.47 & 0.32 & 1.86 & 1.21 \\
    & S-Surface  & \textbf{0.11} & \textbf{0.05} & \textbf{1.52} & \textbf{0.94} \\
\midrule
\multirow{3}{*}{RL (Baseline)} 
    & SMC        & 0.27 & 0.16 & 2.12 & 1.43 \\
    & PID        & 0.58 & 0.42 & 2.11 & 1.37 \\
    & S-Surface  & 0.10 & 0.05 & 1.70 & 1.13 \\
\bottomrule
\end{tabular}
\end{table*}

To further evaluate action selection quality within the proposed framework, we compare six action strategies: the optimal action (OA) selected according to the Lyapunov critic in RL (LAC), four diffusion-generated candidate actions (CA1--CA4), and the baseline RL (LAC) action. As shown in Fig.~8, the episode returns exhibit a clear monotonic degradation from OA to the lower-ranked candidates. Specifically, OA achieves the highest return at $-4620.9 \times 10^{3}$, followed by CA1 ($-4931.1 \times 10^{3}$), RL (LAC) itself ($-5193.4 \times 10^{3}$), CA2 ($-5327.7 \times 10^{3}$), and CA3 ($-5665.0 \times 10^{3}$), while CA4 performs the worst, with a significantly lower return of $-8330.8 \times 10^{3}$. This ordering reflects the effectiveness of the Lyapunov-based evaluation: actions that better satisfy the Lyapunov decrease condition and promote stable system evolution yield higher episodic returns. The substantial performance gap between OA and lower-ranked candidates (e.g., CA4) highlights the importance of both accurate Lyapunov-based action assessment and the diversity introduced by diffusion-generated candidates. Compared to relying solely on RL (LAC), the proposed framework benefits from a richer action search space, enabling the Lyapunov critic to consistently select more stable and higher-performing actions. These results demonstrate that diffusion-driven candidate generation, when combined with Lyapunov-guided selection, yields more robust and reliable control policies in complex underwater environments.

\begin{figure}[!t]
        \centering
        \includegraphics[width=0.99\linewidth]{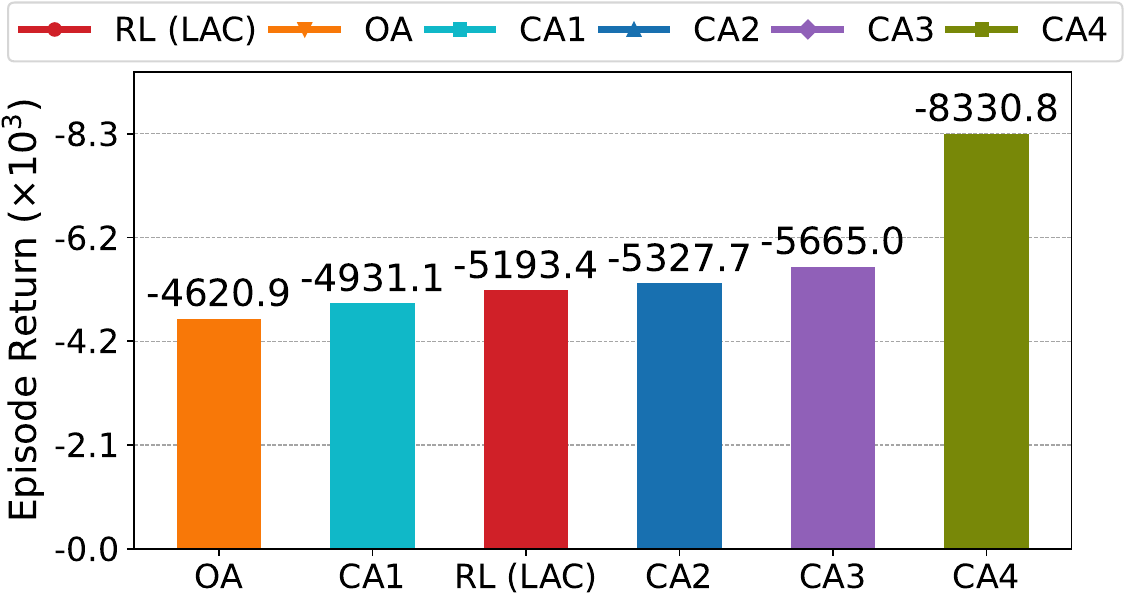}
        \caption{\small Episode returns of six control strategies under the proposed framework, comparing the Lyapunov-guided optimal action (OA), the RL (LAC) baseline, and four diffusion-generated sub-optimal candidate actions (CA1–CA4).}
        \label{fig_5}
\end{figure}

\begin{figure}[!t]
        \centering
        \includegraphics[width=0.99\linewidth]{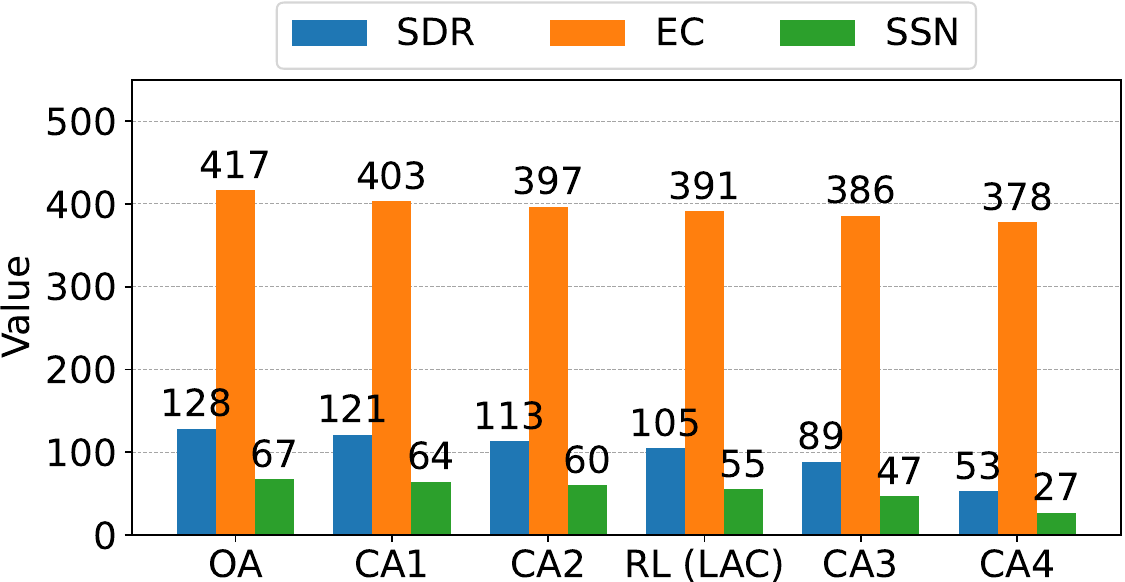}
        \caption{\small Performance comparison across the Lyapunov-guided optimal action (OA), the RL (LAC) baseline, and four diffusion-generated sub-optimal candidate actions (CA1–CA4).}
        \label{fig_9}
\end{figure}

To further assess action quality within the Diffusion+RL framework, we compare six action strategies across a full episode: the optimal action (OA) selected by the Lyapunov critic, four diffusion-generated candidate actions (CA1–CA4), and the baseline RL (LAC) action without diffusion. As shown in Fig.~9, the OA strategy achieves the best task performance across all three metrics, yielding the highest SDR (SDR = 128 MBit/s), the largest number of SSN (SSN = 67), and moderate EC (EC = 417 W). The diffusion-generated candidates show a clear performance trend consistent with their action quality ranking: CA1 and CA2 perform slightly worse than OA (SDR = 121 and 113 MBit/s; SSN = 64 and 60), while still outperforming RL (LAC) (SDR = 105 MBit/s; SSN = 55). In contrast, CA3 and particularly CA4 exhibit strongly degraded performance, with CA4 dropping to only 53 MBit/s in SDR, 27 SSN, and an EC of 378 W—resulting in the lowest overall task quality. These results demonstrate two important insights: (i) diffusion-based candidate generation creates a rich and diverse action set that spans a range of qualities, and (ii) Lyapunov-guided action selection reliably identifies the most stable and high-performing actions within this set. Compared to relying solely on RL (LAC), the proposed framework enables the AUV to execute more informative, energy-efficient, and task-effective trajectories in complex underwater environments.

To enable genuinely task-adaptive stability shaping, our framework leverages an LLM whose reasoning capability directly influences the stability behavior of the learned policy. Rather than depending on manually chosen stability parameters, the LLM semantically interprets task descriptions, environmental conditions, and training feedback, and actively decides how the Lyapunov function and its coefficients should evolve during learning. This LLM-guided process leads to stable behaviors that conventional RL or classical Lyapunov controllers cannot produce. Specifically, the LLM continuously monitors indicators such as transient oscillations, constraint violations, and mission-dependent priorities, then proposes suitable stability weights~$\alpha$ for each stage of training. As shown in Fig.~10, different $\alpha$ values yield visibly distinct contraction patterns of the Lyapunov function within individual episodes. Larger coefficients (e.g., $\alpha = 0.50$) enforce aggressive contraction—rapid initial decay and narrow variance bands—while smaller coefficients (e.g., $\alpha = 0.05$) produce smoother, slower decay. The fact that these behaviors emerge automatically from LLM-driven adjustments highlights its role: the LLM actively balances stability enforcement and control performance, eliminating the need for tedious manual tuning.

Beyond adjusting scalar coefficients, the LLM further contributes by selecting the most appropriate Lyapunov functional form from a pre-defined library, thereby aligning stability shaping with task semantics. Fig.~11 visualizes the effect of three forms—Softplus, Squared, and Log—each exhibiting characteristic contraction curvature. Softplus provides mild stabilization with broader uncertainty; the Squared form enforces strong corrections and yields fast, tight contraction; while the Log form mitigates early instability and relaxes near convergence. These differences show that the structure of the Lyapunov function fundamentally determines stability dynamics. Crucially, our LLM identifies these properties and recommends structural transitions—such as switching from Softplus to a more aggressive Squared form when oscillations arise—achieving a level of context-aware stability modulation that static designs cannot match. Together, these mechanisms demonstrate that LLM-guided Lyapunov shaping is not merely a supplement but a decisive factor in ensuring robust, adaptive, and semantically aligned policy learning.

\begin{figure}[!t]
        \centering
        \includegraphics[width=0.99\linewidth]{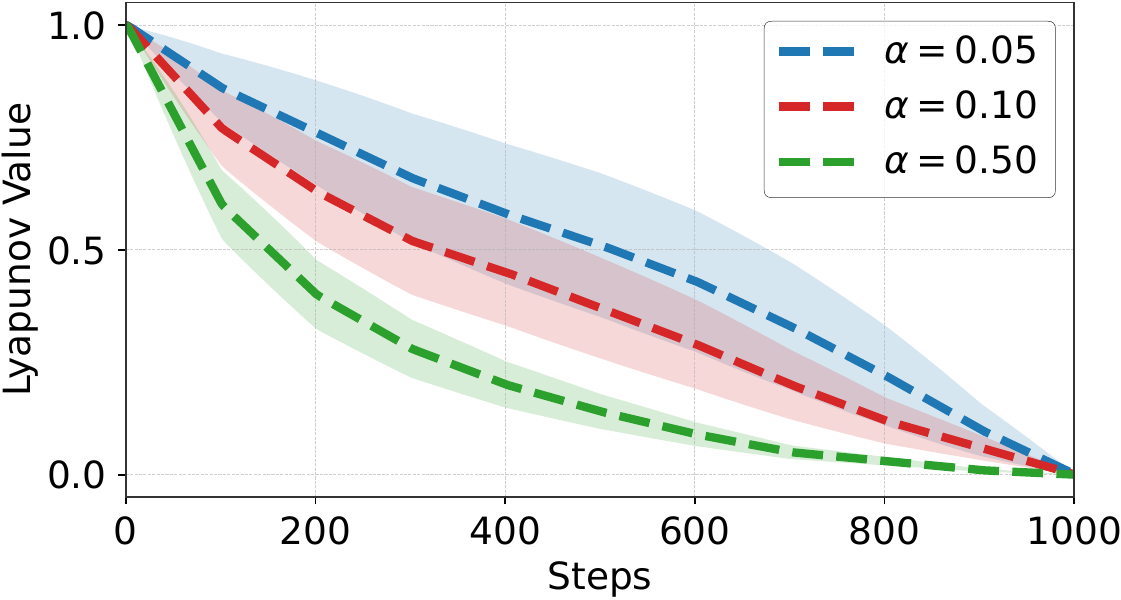}
        \caption{\small Lyapunov value curves under different LLM-selected stability weights $\alpha$, showing monotonic decay and faster, lower-variance contraction for larger $\alpha$, demonstrating task-adaptive stability shaping.}
        \label{fig_7}
\end{figure}

\begin{figure}[!t]
        \centering
        \includegraphics[width=0.99\linewidth]{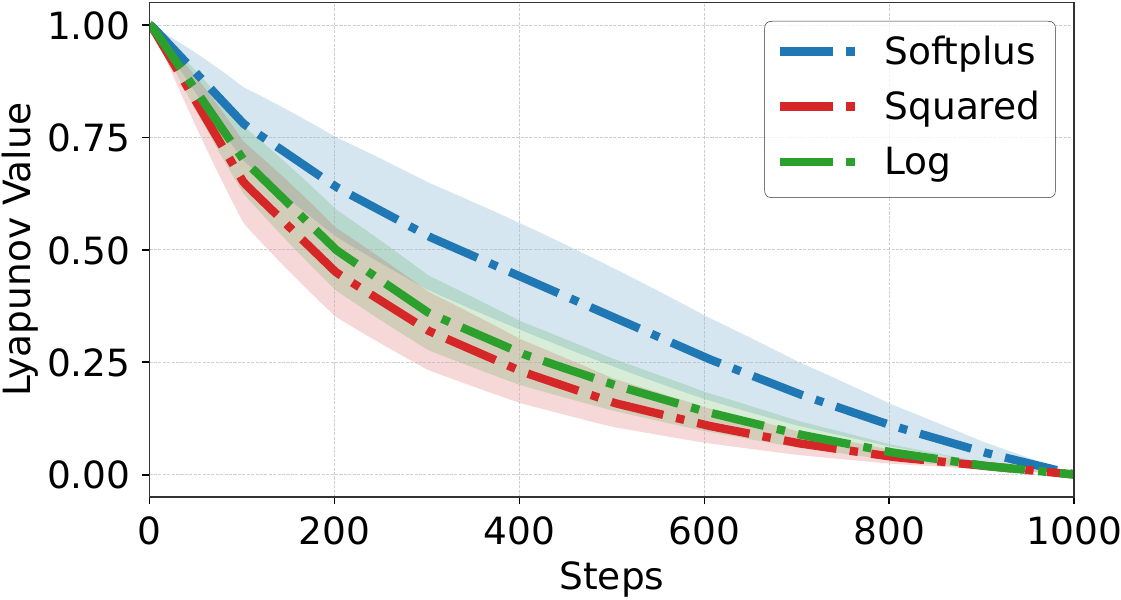}
        \caption{\small Lyapunov value curves under three LLM-selected Lyapunov function forms (Softplus, Squared, and Log), showing distinct contraction behaviors and variance patterns that illustrate how the functional structure influences stability shaping.}
        \label{fig_8}
\end{figure}

\section{CONCLUSIONS}
In this study, we develop a diffusion-prior Lyapunov actor–critic framework that unifies exploration efficiency, stability assurance, and task-level semantic adaptability. This framework integrates a diffusion-based action generator that produces smooth and disturbance-resilient trajectories, a Lyapunov critic that enforces stability and suppresses unreliable behaviors, and an LLM-driven outer loop that adaptively refines Lyapunov functions and parameters based on mission semantics and training feedback. Through this cohesive “generation–filtering–optimization” mechanism, The proposed framework substantially improves sample efficiency, long-horizon planning, and stability-aware policy learning while ensuring theoretical compatibility between diffusion-based exploration and Lyapunov-constrained updates. Extensive simulations under complex ocean dynamics demonstrate that the proposed framework achieves more accurate trajectory tracking, faster convergence, superior energy efficiency, and enhanced robustness compared to both conventional RL and diffusion-augmented baselines, underscoring its promise as a principled and generalizable framework for stable and robust AUV control in underwater tasks.

Looking ahead, several promising directions remain open. First, a deeper theoretical characterization of how the Lyapunov stability weight and different Lyapunov functional forms influence contraction speed, variance propagation, and overall policy optimality would further illuminate the design space of stability-aware RL. Such analyzes may yield principled guidelines for selecting or adapting Lyapunov structures beyond the current LLM-driven heuristics. Second, extending the framework toward real-world deployment—incorporating field-calibrated hydrodynamics, acoustic communication constraints, and onboard computational limitations—will be essential for validating its practicality at sea. Planned pool and coastal experiments with REMUS-class platforms will serve as an intermediate step toward full-scale ocean trials. Together, these efforts aim to advance the proposed from a simulation-verified framework to a robust and deployable solution for real-world AUV autonomy.
    
    \bibliographystyle{IEEEtran}
    \bibliography{TMC-xjzh2}


\end{document}